\documentclass{revtex4-2}
\usepackage{amsmath}
\usepackage{amsfonts}
\usepackage{amssymb}
\usepackage{graphicx}
\usepackage{xcolor}
\begin{document}

\title{Some Remarks on Alternative (or Modified) Theories of Gravity}

	\author{Júlio C. Fabris}
    \email[]{julio.fabris@cosmo-ufes.org}
	\affiliation{Núcleo Cosmo-ufes \& Departamento de Física, Universidade Federal do Espírito Santo, Av. Fernando Ferrari, 514, Goiabeiras, 29060-900, Vit\'oria, ES, Brazil}
    \affiliation{National Research Nuclear University MEPhI (Moscow Engineering Physics Institute), 115409, Kashirskoe shosse 31, Moscow, Russia.}

\begin{abstract}

A seminar given about 30 years ago by Ruben Aldrovandi motivates this text where some reflexions about constructing theories that modify General Relativity are made. Two particular cases, the Brans-Dicke and Unimodular Gravity ones,
are discussed, in a quite qualitative way, showing on how they can address some of the most outstanding problems of General Relativity, specially the transplanckian physics and the cosmological constant problem.

\end{abstract}
	
	
\maketitle

\section{At that time...}

I think it was about 1993, in Paris, in the {\it Laboratoire de Gravitation et Cosmologie Relativistes}, at that time under the direction of Richard Kerner. Ruben Aldrovandi gave a seminar, that was held at the library of this research unity of the {\it Université Pierre et Marie Curie} devoted to gravitation, cosmology and other branches of theoretical physical.
Aldrovandi began the seminar by asking to the audience how many alternatives have been proposed to General Relativity since the formulation of this geometric theory of gravity. Since nobody tried to answer, he gave an approximative estimation: {\it more than one thousand}. Later, during his seminar he tried to develop a possible mechanism for avoiding the initial singularity but keeping the accepted framework of the GR and Elementary Particle Physics. At the end, interesting results were displayed, based on the extreme compression of matter, that needed as expected some new ingredients (but not so exotic in principle) to give a complete scenario. Perhaps Ruben wanted to alert that very good results concerning open and challenging problems could be obtained if
we keep ourselves in the framework of the known physics, with some extrapolations to extreme situations: in facing a challenge, perhaps no need to throw immediately away the physics we have developed for decades, even if an open view must always be welcome. 

Of course, the remark made by Rubem Aldrovandi in his speech in the beginning of the 90's had not the intention to criticize 
the construction of alternatives to GR.  Later, he had made extensive studies in, for exemple, teleparell gravity, an alternative to GR: the gravitational effects in teleparell gravity are given by torsion instead of curvature of  a Riemannian manifold. In an article, written by Ruben Aldrovandi and José Geraldo Pereira, published in 2016, the authors write \cite{rubena}: {\it This new theory is fully equivalent to general
relativity in what concerns physical results, but is deeply different from the conceptual point of view. Its characteristics make of teleparallel gravity an appealing
theory, which provides an entirely new way to think the gravitational interaction}. 

Teleparalell gravity (see Ref. \cite{rubenb} for a detailed description of this theoretical proposal) may give some new insights for the shortcomings of GR, as the difficulties connected with quantization of the gravitational interaction, preserving otherwise the main achievements of the standard theory of gravity. I understood his remark at that time about the huge number of alternative theories of gravity as an indication that, when studying one or another the important open problem, for example in gravitational physics, one must keep in mind the actual motivations to consider GR as the Standard Theory of Gravity and also the reasons to look for alternatives to this theory.

GR is in principle a very successful theory for the gravitational interaction. It explains de orbits of planets and stars, the deviation of light by massive objects, it predicts the existence of gravitational waves and black holes, besides other compacts objects. However, it contains singularities either inside black holes and at the beginning of the universe. In cosmology, the success in the predictions concerning the primordial nucleosynthesis and of a hot phase in the history of the universe are remarkable. However, without postulating a dark sector for the universe (dark matter and dark energy), it is not possible to understand the present stage of accelerated expansion of the universe, the dynamics of virialized cosmic system and even the formation of the structures in the observed cosmos. All tentative to detect directly the components components of the dark sector have failed until now.

From theoretical side, GR is not easy to quantize. Even if much progress has been obtained in the direction of the construction a consistent quantum theory of gravity, it is clear that to quantize the space-time itself is not an easy task. Teleparalellism may perhaps give a good new path
in order to construct a quantum theory of gravity. But, it is still until now an interesting possibility: a concrete quantum theory of gravity has not yet be presented using Teleparalell theory of gravity. Hence, there are reasons to consider viable alternatives to the Standard Theory of Gravity. But, it is important not to loose a general view over the overall scenario, and mainly the success and
difficulties of the current Standard Theory of Gravity.

Here, I will make some considerations on two possible alternative theories of gravity, the Brans-Dicke theory (with some unavoidable extension) and the Unimodular Gravity. It is interesting to remember that {\it alternative theories of gravity} looks an old fashion terminology, perhaps consequence of the social ambiance of the 60's (mainly) and 70's, when contestation of the occidental standard society proliferated from different sides. Today, {\it modified gravity} seems to be preferred, perhaps because it became important to stress that GR is a consistent theory of gravity, even if there are difficulties here and there. I think that the Brans-Dicke and Unimodular frameworks may be good examples on the different possibilities to construct a theory of the gravitational interaction outside the strict GR context, addressing some important open issues. We will try to show our point of view for this. The focus will be on a qualitative, and incomplete, discussion on the possibilities that are opened when some of the cornerstones of GR are abandoned in favor of a non standard framework.

\section{The variation of the fundamental constants}

The laws of physics make appeal to some characteristics, fundamental constants. These constants simbolizes the domain of physics involved in a given expression. For example, the gravitational coupling constant $G$ indicates that gravitation is, at least, one of the phenomena under consideration. The presence, of $\hbar$ indicates that we have to do with quantum mechanics, the presence of $c$ (velocity of light) to relativistic domain, the Boltzmann constant $k_B$ with thermodynamics. In the General Relativity equations,
\begin{eqnarray}
R_{\mu\nu} - \frac{1}{2}g_{\mu\nu}R = \frac{8\pi G}{c^4} T_{\mu\nu},
\end{eqnarray}
it appears explicitly $G$ and $c$. Hence, it is relativistic theory of gravity, while in the Newton's theory of gravity, expressed by the 
Poisson equation,
\begin{eqnarray}
\nabla^2\Psi = 4\pi G\rho,
\end{eqnarray}
for the gravitational potential $\Psi$,
only $G$ appears: it is a non-relativistic theory of gravity.

To my knowledge, the only physical expression that contains all these constants is the black hole temperature. For the Schwarzschild black hole, the temperature reads,
\begin{eqnarray}
T = \frac{\hbar c^3}{8\pi G M k_B}.
\end{eqnarray}

We can add to the list of these fundamental constants the electric charge, $e$. It denotes the presence of electromagnetic interaction. The observations in astrophysics and cosmology are made through the detection of electromagnetic radiation, and in this sense it is important to take into account $e$ in our description. In fact, it appears in the fine structure constant,
\begin{eqnarray}
\alpha = \frac{e^2}{\hbar c} \approx \frac{1}{137}.
\end{eqnarray}

From $\hbar$, $G$ and $c$, we can construct characteristic unities of length, time and mass:
\begin{eqnarray}
L_P &=& \sqrt{\frac{G\hbar}{c^3}} \sim 2\times10^{-33}\,\mbox{cm},\\
T_P &=& \sqrt{\frac{c^5}{G\hbar}} \sim 5\times10^{-44}\,\mbox{s},\\
M_P &=& \sqrt{\frac{\hbar c}{G}} \sim 2\times10^{-5}\,\mbox{g}.
\end{eqnarray}
From these expression, we can obtain the Planck energy,
\begin{eqnarray}
E_P = \sqrt{\frac{\hbar c^5}{G}} \sim 2\times10^{16}\,\mbox{erg} \sim 1,9\times10^{19}\,\mbox{GeV},
\end{eqnarray}
and the Planck temperature,
\begin{eqnarray}
T_P = \sqrt{\frac{\hbar c^5}{Gk_B^2}} \sim 10^{32}\,\mbox{K}.
\end{eqnarray}

These expressions contains the constant $G$, $\hbar$ and $c$ and, in this sense, they are considered as characteristics
of the regime of a quantum theory of gravity. Their magnitude indicate that we are very far from a direct test of the quantum gravity regime:
The maximum energy produced in laboratory (LHC) is about $10$\,TeV, testing a distance of about $2\times10^{-18}$\,cm, with a time scale of $10^{-28}$\,s, and temperature of the order of $10^{17}$\,K. However, some indirect tests of the quantum gravitational regime may be considered, as the power spectrum of scalar perturbations imprinted in the large scale structure of the universe, as well as the power spectrum of the primordial gravitational waves (undetected until now, but with good possibilities to be detected in the near future). The primordial spectrum is determined by the microphysics in the very early universe. Hence, a quantum gravitational phase may leave traces in the primordial spectrum. But the exact determination of this fossils of a quantum gravitational era depends on the specific theory of quantum gravity, which is yet under construction. For some tentatives of obtaining a primordial spectrum in the context of a string effective theory, for example, see Refs. \cite{gasperini,npf}.

In the 30's, Dirac considered the hypothesis that the fundamental constants may, in fact, vary with time \cite{dirac}. The reasoning of Dirac is based on the large number hypothesis, identifying some coincidences in specific numbers obtained with fundamental constants. For example, using $c$, $G$ and the Hubble constant $H_0$, we can recover the pion mass:
\begin{eqnarray}
\label{coin}
\biggr(\frac{\hbar^2 H_0}{Gc}\biggl)^\frac{1}{3} \sim m_\pi.
\end{eqnarray}
This relation appeared very important at the time Dirac exposed his speculations since the studies of the atomic nucleus was taking form 
and the strong interaction has been proposed in order to retain the nucleus stable. The strong interaction should be mediated by a massive particle, identified as the pion, with a given mass ($140$\,MeV) corresponding to a typical range for this interaction of the order of the size of the atomic nucleus.
In order the relation above not to be a pure coincidence but reflecting a deeper law of Nature, and since the Hubble constant is connected with the age of the universe, $c$ and/or $G$ should depend on time. The choice made by Dirac was that $G$ must be a function of time, such that,
\begin{eqnarray}
G =  G_0\frac{t_0}{t},
\end{eqnarray}
where $t_0$ is the present time and $G_0$ the present value of the gravitational coupling.

Another possibility, it would be to take the velocity of light as depending on time. This hypothesis led to many new proposals connected with a violation of Lorentz symmetry. There are strong constraints on a varying speed of light and, consequently, on the violation of Lorentz symmetry \cite{liberati,lee}. However, this hypothesis is still not excluded. 

Concerning the variation of the gravitational coupling term $G$, there are very strong observational constraints. The most recent one indicates that the fractional allowed variation of $G$ is of the order of \cite{G},
\begin{eqnarray}
\frac{\dot G}{G} \leq 10^{-13}\,\mbox{year}^{-1}.
\end{eqnarray}
Since the age of the universe is $T_U \sim 10^{10}$\,year, we can, very crudely estimate that the $G$ has varied about
a fraction of $10^{-3}$ during all the cosmic history. This would corresponds to a change in the third decimal case in its value. It is not enough to bring the Planck energy, for example, for a significative lower value. 

However, even with this strong observational constraint, the possibility that $G$ is a dynamical quantity may allow some very interesting results: by extrapolating to
times near the initial singularity, the variation of $G$ can be more important with far reaching consequences. We will describe in next section how to implement a variation of $G$ in a relativistic theory of gravity, in occurrence the Brans-Dicke theory, and some of its possible consequences, mainly regarding a hypothetical new window to the Planck era and some different views of the Standard Cosmological Model, in particular the expansion of the universe.

\section{Scalar tensor theories: some curiosities in the Brans-Dicke case}

The Dirac proposal of considering $G$ as a function of time has not been followed by a concrete theory incorporating this idea. Further steps to implement a theory of gravity may be traced back to Jordan, at the end of the 40's in the last century, also in an article in the journal Nature \cite{jordan}. Jordan considered also in this work the possibility of matter creation in an expanding universe.

A consistent, relativistic theory of gravity with a dynamical gravitational coupling has been proposed in the beginning
of the 60's by C. Brans and R.H. Dicke, the now called Brans-Dicke (BD) theory \cite{bd}. In the BD theory, $G$ is replaced by a dynamical, long interaction scalar field $\phi$, such that,
\begin{eqnarray}
G \propto \frac{1}{\phi}.
\end{eqnarray}
A kinetic term for the scalar field is introduced together with a dimensionless coupling parameter $\omega$. No potential related to $\phi$ term is added, in conformity with the proposal that $\phi$ is a long interaction scalar field: the presence of a potential term, describing a self-interaction, would lead to an effective mass, shortening the range of the interaction.

In this way, the Einstein-Hilbert Lagrangian, 
\begin{eqnarray}
{\cal L} = \sqrt{-g}\biggr\{\frac{R}{16\pi G}\biggl\} + {\cal L}_m(g_{\mu\nu},\Psi),
\end{eqnarray}
with the matter Lagrangian given by ${\cal L}_m(g_{\mu\nu},\Psi)$, $\Psi$ representing the matter fields, is replaced by
\begin{eqnarray}
\label{bd-action}
{\cal L} = \sqrt{-g}\biggr\{\phi R - \omega\frac{\phi_{;\rho}\phi^{;\rho}}{\phi}\biggl\} + {\cal L}_m(g_{\mu\nu},\Psi).
\end{eqnarray}

The GR limit must be obtained by imposing $\phi$ constant but also by considering $\omega \rightarrow \infty$. However, in some particular situations, this limit is not so well defined \cite{cr,ft}. The observational constraints indicate a very large value for $\omega$, up to $\omega > 40,000$ \cite{will}. In this case, in principle the theory must be essentially indistinguishable from GR. But, we must stress the expression `in principle".

From the theoretical side, there are some interesting connections of the BD theory with other, perhaps more fundamental, theories. For example, at low energy, the effective action emerging from the string theory, is given by \cite{cw},
\begin{eqnarray}
{\cal L} = \sqrt{-g}\biggr\{\phi R + \frac{\phi_{;\rho}\phi^{;\rho}}{\phi}\biggl\},
\end{eqnarray}
neglecting matter and gauge fields.
This is the BD theory with $\omega = - 1$. Of course, this value of $\omega$ is in strong disagreement with observations, but we will return to this question below. 
Also, multidimensional theories, à la Kaluza-Klein, may lead, after reduction to four dimensions, to a BR type action, with
\begin{eqnarray}
\omega = - \frac{(d - 1)}{d},
\end{eqnarray}
$d$ being the number of extra dimensions, $D = 4 + d$.
Hence, KK theories \cite{bl} with arbitrary dimension may have connection with BD theory with $\omega$ varying from $0$ ($D = 5$) and $- 1$, which coincides with the effective string action, but under the condition $D \rightarrow \infty$.
We must quote also that the, today very fashion, $f(R)$ theories may be mapped in a BD-type theory, with $\omega = 0$, but with a potential determined by the form of function $f(R)$ \cite{tf}. The presence of the potential, in this case, is in strong contrast with the original purposes of the BD theory, however.

For the moment, let us ignore the observational constraints on $\omega$, and let us explore some aspects of the Brans-Dicke theory. We write first the field equations from the action. They read,
\begin{eqnarray}
\label{fe1}
R_{\mu\nu} - \frac{1}{2} g_{\mu\nu}R &=& \frac{8\pi}{\phi}T_{\mu\nu} + \frac{\omega}{\phi^2}\biggr(\phi_{;\mu}\phi_{;\nu} - \frac{1}{2}g_{\mu\nu}\phi_{;\rho}\phi^{;\rho}\biggl) + \frac{1}{\phi}\biggr(\phi_{;\mu}\phi_{;\nu} - g_{\mu\nu}\Box\phi\biggl),\\
\label{fe2}
\Box\phi &=& \frac{8\pi T}{3 + 2\omega},\\
\label{fe3}
{T^{\mu\nu}}_{;\mu} &=& 0.
\end{eqnarray}
First of all, a remark on equation (\ref{fe3}): as in GR, in the BD theory the energy-momentum tensor is conserved since both theories are invariant by general diffeomorphic transformation \cite{wald}, a property that will be discussed in more details below.

We will now specialize the field equations to the flat Friedmann-Lemaître-Robertson-Walker (FLRW) metric, given by,
\begin{eqnarray}
ds^2 = dt^2 - a(t)^2(dx^2 + dy^2 + dz^2).
\end{eqnarray}
The resulting equations of motion are the following:
\begin{eqnarray}
3\biggr(\frac{\dot a}{a}\biggl)^2 &=& \frac{8\pi}{\phi}\rho + \frac{\omega}{2}\frac{\dot \phi^2}{\phi^2} - 3H\frac{\dot\phi}{\phi}, \\
\ddot \phi + 3 H\dot\phi &=& \frac{8\pi}{3 + 2\omega}(\rho - 3p),\\
\dot\rho + 3H(\rho + p) &=& 0.
\end{eqnarray}

We consider two particular solutions for these equations according the equation of state connecting $p$ and $\rho$.
\begin{itemize}
\item Vacuum state, $p = - \rho$:
\begin{eqnarray}
a &\propto& t^{\omega + 1/2},\\
\phi &\propto& t^2;
\end{eqnarray}
\item Incoherent matter, $p = 0$:
\begin{eqnarray}
a &\propto& t^\frac{2 + 2\omega}{4 + 3\omega},\\
\phi &\propto& t^\frac{2}{4 + 3\omega}.
\end{eqnarray}
\end{itemize}

The solutions above admit a constant scale factor $a$. For the first case, it occurs for $\omega = - 1/2$, while for the second case, this configuration occurs for $\omega = - 1$, a value that coincides with the string low energy limit.
This does not mean properly a static universe, since the gravitational coupling $\phi$ is a function of time.
In both cases, $\phi \propto t^2$, and the gravitational coupling diverges as $t \rightarrow 0$. Hence, in these cases, the Planck energy goes to zero asymptotically. This property may have important consequences for the physics in the primordial universe, affecting for example the features of the initial quantum fluctuations responsible for the primordial spectrum of scalar and tensorial perturbations.

A remark to be made is that these solutions with a constant scale factor are stable as demonstrated in Ref. \cite{plinio}. For example, for $p = - \rho$, the perturbations on the scalar field, expressed as
\begin{eqnarray}
\lambda = \frac{\delta\phi}{\phi},
\end{eqnarray}
$\delta\phi$ being the fluctuation on the (inverse) gravitational coupling, and using the synchronous coordinate condition,
reads \cite{plinio},
\begin{eqnarray}
\lambda = \frac{1}{t}\biggr\{\int t^\frac{1 - 5r}{2}\biggr[C_1 J_k(y) + C_2J_{-k}(y)\biggl]dy + C_3\biggl\},
\end{eqnarray}
with the definitions,
\begin{eqnarray}
x = t^p, \quad p = 1 - r, \quad r = \omega + \frac{1}{2}, \quad y = \frac{kx}{1 - r}, \quad q = \frac{4r + 3}{2(1 - r)},
\end{eqnarray}
$k$ being the wavenumber and $C_i$ integrations constants.

For the static scale factor in the case of a vacuum fluid ($p = - \rho$, $\omega = - 1/2$), the solution reads (neglecting the fictitious solution represented by the constant $C_3$),
\begin{eqnarray}
\lambda = \frac{1}{t}\biggr\{\int t^\frac{1}{2}\biggr[C_1 J_{3/2}(k\,t) + C_2J_{-3/2}(k\,t)\biggl]dx\biggl\}.
\end{eqnarray}
The {\it static} solution is stable, as it can be verified, for example, by inspecting the asymptotical behavior. The same happens with the static case with $p = 0$, and $\omega = - 1$. We must remember that a solution with constant scale factor in GR is unstable. 

Of course, we deal here with a fictitious static universe. Even if the scale factor is constant, the gravitational coupling varies. In both cases, $p = - \rho$ and $p = 0$, it varies as $G \propto t^{-2}$. It means that $G$ is a decreasing function of time. Hence, the Planck energy was smaller in the past, approaching zero as $t \rightarrow 0$. 
The time elapsed from this moment until today is finite. In principle, we can have two disconnected branches, one with
$- \infty < t < 0$, and another with $0 < t < \infty$. If just one branch is consider, there is no curvature singularities (since $a$ is constant), but there is a geodesic singularity. A bounce-type scenario must be obtained in order to get rid off this geodesic singularity.

Of course, at this level, the above description is just a curiosity. But, it can be brought to a more serious scenario. For example, how to explain the cosmological redshift? We remark that in the gravitational interaction what matters frequently is $GM$, and the variation of $G$ can be translated into a variation of mass, leading to a change in the spectral lines.

However, there must be a mechanism in order to pass from one cosmological phase to another. We observe a relic of a hot phase, the Cosmic Microwave Background radiation, indicating a radiative dominated phase. Besides this, the observed abundance of helium requires also a hot phase in the past, the existence of structures like galaxies requires the presence of matter with an effective pressure near zero, etc. In order, to implement all these observations in such a scenario, we need to introduce a new ingredient. For example, to pass from a vacuum phase to
a matter phase, but keeping the scale factor constant, it is required to change the value of $\omega$ from $- 1/2$ to $- 1$. This can be achieved by considering
$\omega$ as a function of time. It is, in some sense, an {\it epicycle}, but not completely out of sense. Let us describe how it can be obtained. 

Our start point is the Bergmann-Wagoner-Brans-Dicke theory whose Lagrangian in the original Jordan frame is \cite{bd,w}
\begin{eqnarray}
{\cal L} = \sqrt{-g}\biggr\{\phi R - \omega(\phi)\frac{\phi_{;\rho}\phi^{;\rho}}{\phi}\biggl\} + {\cal L}_m(g_{\mu\nu},\Psi),
\end{eqnarray}
with, as before, the matter Lagrangian given by ${\cal L}_m(g_{\mu\nu},\Psi)$, $\Psi$ representing the matter fields. The novelty is that $\omega$ now is a function of $\phi$.

The field equations are the following:
\begin{eqnarray}
R_{\mu\nu} - \frac{1}{2} g_{\mu\nu}R &=& \frac{8\pi}{\phi}T_{\mu\nu} + \frac{\omega(\phi)}{\phi^2}\biggr(\phi_{;\mu}\phi_{;\nu} - \frac{1}{2}g_{\mu\nu}\phi_{;\rho}\phi^{;\rho}\biggl) + \frac{1}{\phi}\biggr(\phi_{;\mu}\phi_{;\nu} - g_{\mu\nu}\Box\phi\biggl),\\
\Box\phi &=& \frac{8\pi T}{3 + 2\omega(\phi)} - \frac{\omega_\phi}{3 + 2\omega(\phi)}\phi_{;\rho}\phi^{;\rho},\\
{T^{\mu\nu}}_{;\mu} &=& 0.
\end{eqnarray}
The subscript in $\omega$ indicates derivative with respect to $\phi$.

Since $\omega$ is now a function of the scalar field, $\phi$, it can change its value, for example, from $- 1/2$ to $-1$. For this we can suppose that,
\begin{eqnarray}
\omega(\phi) = - \frac{1}{2}\tanh\phi - \frac{1}{2}.
\end{eqnarray}
The solutions may differ from those display above, admitting for example $\phi \propto t$, recovering the original proposal of Dirac,
$G \propto \frac{1}{t}$.  The main point is that the range of scalar field must be $0 < \phi < \infty$.
There is one more difficulty to be surmounted: to pass from one phase, dominated by a given fluid to another one, dominated by another fluid, it is necessary that the fluid may depend on $\phi$. This can not be achieved here, since the conservation of the energy-momentum tensor, with a constant, scale factor, implies $\rho_i = $ constant for all matter components. At this stage, we can circumvent this problem by rewritting the field equations in the minimally coupled framework through a conformal transformation, such that,
\begin{eqnarray}
g_{\mu\nu} = \phi^{-1}\tilde g_{\mu\nu}.
\end{eqnarray}
In this case, the equations read now,
\begin{eqnarray}
\tilde R_{\mu\nu} - \frac{1}{2} \tilde g_{\mu\nu}\tilde R &=& 8\pi G\tilde T_{\mu\nu} + \frac{\frac{3}{2} + \omega(\phi)}{\phi^2}\biggr(\phi_{;\mu}\phi_{;\nu} - \frac{1}{2}\tilde g_{\mu\nu}\phi_{;\rho}\phi^{;\rho}\biggl),\\
\tilde \Box\phi &=& \frac{8\pi G\phi T}{3 + 2\omega(\phi)} - \biggr(\frac{\phi \omega_\phi }{3 + 2\omega(\phi)} - 1\biggl)\frac{\phi_{;\rho}\phi^{;\rho}}{\phi},\\
{\tilde T^{\mu\nu}}_{;\mu} &=& -\frac{1}{2}\frac{\phi^{;\nu}}{\phi}\tilde T.
\end{eqnarray}
In these equations, $G$ is the present gravitational coupling and,
\begin{eqnarray}
\tilde T^{\mu\nu} = (\tilde\rho + \tilde p) \tilde u^\mu \tilde u^\nu - \tilde g^{\mu\nu}\tilde p,
\end{eqnarray}
with,
\begin{eqnarray}
\tilde\rho = \frac{\rho}{\phi^2}, \quad \tilde p = \frac{p}{\phi^2}.
\end{eqnarray}
In this case it is possible to transit from a phase dominated by a given fluid to another phase dominated by another fluid. But, in order to do this the constancy of the scale factor must be implement in the minimal coupled frame described above, while in the original frame the universe would be fully dynamical, with a varying scale factor: a fully dynamical universe, can be mapped in a universe without expansion.

The reasoning exposed above reveals that it is very challenging to take very far the {\it curiosity} of having a static scale factor $a$ in a stable cosmological scenario, trying to construct a realistic and complete model. But, at this stage this is not in principle excluded.
If this program is, at least partially, successful it may bring new visions on the observational facts like the cosmological redshift, dark energy, initial singularities, etc. Our description above has some similarities (and also some important differences) with the scenario of a universe without (or almost without) expansion proposed by Wetterich \cite{wette}. What we have described above is that a theory like the Brans-Dicke may also admit a universe without expansion, even if facing many obstacles to construct a realistic cosmological scenario.

In any case, it may be stressed that, even in its traditional formulation, BD theory may open a window to investigate the transplanckian regime.

\section{The cosmological constant: a problem?}

In the previous section, we have discussed some curious possibilities that emerge from considering the gravitational coupling $G$ as a dynamical quantity. In particular, how to obtain a universe "without expansion" has been considered within one of the most important relativistic theory that incorporates a variable gravitational coupling in a relativistic context, the Brans-Dicke theory. Now, we turn to another recurrent topic in modern cosmology, the presence of the cosmological constant in cosmological models coming from GR. This is frequently considered as one of the most important problem in 
physics today. But, to which extend this is a problem? We will try to address, at least partially, this question here.

The so-called {\it cosmological constant problem} has a long history. The first frequently quoted discussion comes from the attempt to construct a static model for the universe by A. Einstein. In cosmology, using the non-flat FLRW metric and a pressureless matter component, the equations governing the evolution of the universe are given by,
\begin{eqnarray}
H^2 + \frac{k}{a^2} &=& \frac{8\pi}{3}G\rho, \\
2\dot H + 3H^2 + \frac{k}{a^2} &=& 0,\\
\dot\rho + 3H\rho &=& 0.
\end{eqnarray}
For a static universe ($H = 0$), these equations are only consistent with $k = 0$ and $\rho = 0$, that is, an empty Minkowski space-time. Inserting a cosmological term, the equations become,
\begin{eqnarray}
H^2 + \frac{k}{a^2} &=& \frac{8\pi}{3}G\rho + \frac{\Lambda}{3}, \\
2\dot H + 3H^2 + \frac{k}{a^2} &=& \Lambda,\\
\dot\rho + 3H\rho &=& 0.
\end{eqnarray}
If the universe is flat ($k = 0$), again there is only trivial solution, but for $k \neq 0$, the solutions become,
\begin{eqnarray}
\frac{2}{3}\Lambda &=& \frac{8\pi}{3}G\rho, \\
\frac{k}{a^2} &=& \Lambda.
\end{eqnarray}
For $k < 0$ (open universe), a static universe requires negative mass, but for closed universe ($k > 0$), a positive mass, with a positive cosmological constant, leads to the desired static configuration. Moreover, the universe will have a spherical geometry. Not bad, but this universe is unstable, as it was quickly realized.

This negative result for a static universe does not led to the complete rejection of the cosmological term. In the 30's, Lemaître has revived the cosmological constant for another reasons \cite{lm}. At that moment, the density of the universe were estimated (very crudely) by the distribution of galaxies, let us say, only by the detected baryonic matter (The possibility of the existence of a {\it dark } matter component was only evoked by Zwicky, with very unprecise estimations \cite{z}).
After the works of Friedmann, Lemaître and Hubble, the possibility of a static universe has been rejected in favor of dynamical, expanding universe. Using the present data, the baryonic density today is of the order of \cite{planck},
\begin{eqnarray}
\rho_{b0} \sim 3,6\times10^{-28}\,\frac{\mbox{kg}}{\mbox{m}^3} = 3,6\times10^{-31}\,\frac{\mbox{g}}{\mbox{cm}^3}.
\end{eqnarray}
The subscript $0$ indicates the present value.

However, the estimations of the baryonic component were very crude in the 30's, and it was considered the possibility of a universe dominated by baryons only, leading to the Einstein-de Sitter universe.
Hence, we can consider the Friedmann's equation,
\begin{eqnarray}
H^2 = \frac{8\pi G}{3}\rho_{b}.
\end{eqnarray}
Using that $\rho_b = \rho_{b0}a^{-3}$
we obtain,
\begin{eqnarray}
a = a_0\biggr(\frac{t}{t_0}\biggl)^{2/3} .
\end{eqnarray}
The age of the universe, today, is given by,
\begin{eqnarray}
t_0 = \frac{2}{3}\frac{1}{H_0}.
\end{eqnarray}
Using the measured value of $H_0$, for example obtained from the Cepheid stars,
\begin{eqnarray}
H_0 \sim 70 \frac{\mbox{km}}{\mbox{Mpc}\cdot\mbox{s}},
\end{eqnarray}
we obtain
an age of the universe of the order of $9,5$ billion years, smaller than the age of the globular clusters (the oldest virialized stellar system known), which is of the order of the 13 billion years.

Everything changes if a cosmological constant is added. In this case, the Friedmann equation becomes,
\begin{eqnarray}
H^2 = \frac{8\pi G}{3}\rho + \frac{\Lambda}{3},
\end{eqnarray}
Since $\Lambda$ is a constant, this equation admits the solution (supposing the matter component with the same behavior as before), given by,
\begin{eqnarray}
a(t) = \biggr(\frac{\Omega_m}{\Omega_\Lambda}\biggl)^{1/3}\sinh^ {2/3} \biggr\{\frac{3}{2}\sqrt{\Omega_\Lambda}t\,H_0\biggl\},
\end{eqnarray}
with 
\begin{eqnarray}
\Omega_{m} &=& \frac{8\pi G}{3H_0^2},\\
\Omega_\Lambda &=& \frac{\Lambda}{3H_0^2},
\end{eqnarray}
with $\Omega_m + \Omega_\Lambda = 1$.

The Hubble function is given by,
\begin{eqnarray}
H_0 = \frac{\dot a}{a}\bigg|_{t = t_0}
\end{eqnarray}
where $t_0$ is the present time. Solving numerically the resulting relation with, for example, $\Omega_\Lambda = 0.7$, we find
$t_0 = 13.1$ billion years. This result differs from present estimations (13.8 billion years) because we are ignoring the complete evolution of the. universe. But, this simple analysis show how the introduction of the cosmological constant may solve the {\it age crisis}.

Hence, the presence of the cosmological constant in the Friedmann equation can solve one of the main problems of the old standard cosmological model based only on the baryonic matter: the incompatibility between the age of the universe and the age of some structures that emerge in the course of the evolution of the cosmos, in occurrence, the globular clusters, as pointed out by Lemaître. Moreover, the introduction of the cosmological constant is very natural. As it was established much later, in the 70's, by Lovelock \cite{love}, the most general action constructed out from the curvature tensor, leading to second order differential equations, is given by,
\begin{eqnarray}
I_D = \sum_{n = 0}^{[D/2]}c_n\frac{1}{n!}\delta^{[\alpha_1\alpha_2\cdot\cdot\cdot\alpha_{n-1}\alpha_n]}_{[\beta_1\beta_2\cdot\cdot\cdot\beta_{n-1}\beta_n]}{R^{\beta_1\beta_2}}_{\alpha_1\alpha_2}\cdot\cdot\cdot{R^{\beta_{n-1}\beta_n}}_{\alpha_n\alpha_{n-1}}.
\end{eqnarray}
In this expression $D$ is the dimension of space-time, $[D/2]$ indicates the integer part, and anti-symmetrization 
in the indices $\alpha$ and $\beta$ are indicated, while $c_n$ are arbitrary constants.
In four dimensions
we find,
\begin{eqnarray}
I_4 = \Lambda + c_1 R + c_2\biggr(R_{\mu\nu\rho\sigma}R^{\mu\nu\rho\sigma} - 4 R_{\mu\nu}R^{\mu\nu} + R^2\biggl).
\end{eqnarray}
In this expression, $c_0$ has been identified with the cosmological constant, $R$ leads to the Einstein-Hilbert Lagrangian, and the term connected with $c_2$ is the Gauss-Bonnet action, which does not contribute to the equation of motions in four dimensions, being a topological term.

The Lovelock invariants reveal that the introduction of the cosmological constant in the gravitational equations are natural. Perhaps, unnatural it would be to ignore it! Of course, it could be argued that nothing predicts the observed value of cosmological constant. But, nothing also previews the value of the gravitational coupling $G$ or other fundamental constant of Nature.
Perhaps a complete and consistent quantum theory of gravity may explain this value. However, we are not still there.

Where it is the {\it cosmological constant problem}? An article of 1989 by Weinberg \cite{wei-cc} discusses extensively this point. Here, we present our point of view. In an article dated of 1965 \cite{gliner}, Gliner argued that the quantum vacuum should have an
equation of state $p = - \rho$, the same equation of state of the cosmological constant, if we interpret the cosmological constant as a fluid. The argument of Gliner is that only this equation of state is invariant by Lorentz transformation, and all inertial observers must {\it see} the same vacuum state.

In fact, suppose a energy-moment tensor on the form of a perfect fluid,
\begin{eqnarray}
T_{\mu\nu} = (\rho, p, p, p).
\end{eqnarray}
Now perform a Lorentz transformation and impose that the transformed energy-momentum tensor has the same form:
\begin{eqnarray}
T_{\mu'\nu'} = (\rho', p', p', p') = \Lambda_{\mu'}^\rho\Lambda_{\nu'}^\sigma T_{\rho\sigma} = (\rho, p, p, p).
\end{eqnarray}
We may insert the usual Lorentz transformation connecting two inertial observers with a relative motion along the axis $x$, for simplicity.
The solution is $\rho' = - p' =  \rho = - p$, that is, the equation of state for a cosmological constant identified as a fluid.

If we have an interpretation of the cosmological constant as a fluid related to the quantum vacuum state, we end with a huge discrepancy. To explain the present accelerated phase of expansion of the universe, we must have a energy density of this fluid, as
\begin{eqnarray}
\rho_\Lambda \sim 10^{-47}\,\mbox{GeV}^4.
\end{eqnarray}
On the other hand, the theoretical estimations of the vacuum energy density in quantum field theory are quite complex, and perhaps we can not say that we know it for sure, since it depends on many phenomena taking place in this {\it vacuum state}. Weinberg gives, in the quoted article reference, initally a crude, very simplified estimation (before launching a more deep analysis), based on the vacuum state of an harmonic oscillator, with a cut off given by the Planck frequency: 
\begin{eqnarray}
\rho_{vac} = \frac{\hbar}{2}\int_0^{\omega_{Pl}} \omega \frac{d^3\omega}{(2\pi)^3c^3} =  \frac{\hbar}{(2\pi)^2} \frac{\omega_{Pl}^4}{c^3}.
\end{eqnarray}
leading to $\rho_{vac} \sim 10^{71}$\, GeV$ ^4$. This leads to a discrepancy between the theoretical and observational value of almost 120 orders of magnitudes, considered as the worst discrepancy in all domain of physics. More detailed (and possibly incomplete still) computations and the introduction of new ingredients (like supersymmetry) may lead to less dramatic estimations (perhaps 60, 50 orders of magnitude). However, there is a consensus that the discrepancy is large.

An ideia that came is to {\it degravitate} the vacuum energy, such that the gravitational effects the vacuum energy are strongly attenuated or even eliminated \cite{dg1,dg2,dg3}. But, this is a proposal in construction. There are cosmological models based on a self-interacting scalar field, called quintessence \cite{quint}, of tachyonic condensate in string theory \cite{cg1,cg2,cg3,cg4} which give up (in a certain way) the use of the cosmological constant to explain the dark energy. However, these models require a zero (or almost zero) vacuum energy density, quite difficult to explain using the reasoning above. In next section, we discuss another proposal which, in principle, modifies substantially the theory of gravitation, that is, General Relativity, but may give a new view of the problem.

\section{Unimodular Gravity}

General Relativity was proposed as the theory of gravity at the end of 1915. It is based on the idea of identifying the gravitational phenomena as the geometry of the space-time in four dimensions, three spatial and one temporal. It was quickly realized that the new gravitational theory could explain the observed anomalous advanced of the perihelium of Mercury, and in 1919 the predictions for the deflection of light was confirmed during the observations of solar eclipse, made in Sobral (Brazil) and Ilha do Príncipe (Africa). More recently, the detection of gravitational wave and of black holes confirmed two other remarkable predictions of GR. However, these outstanding success of the GR is also followed by many difficulties like the existence of the dark sector in the global content of matter and energy of the universe. The cosmological constant problem discussed in the previous section is one of the aspects of the problem associated to the possible existence of the dark sector.

Soon after the final formulation of the GR theory, one of the first variant (among the more one thousand other proposals that were made since then) of the theory has been proposed. It was called Unimodular Gravity (UG), since it has the same content as GR but with constraint on the determinant of the metric \cite{ein,pereira,rev}. Originally, it has been imposed that,
\begin{eqnarray}
\label{umc}
det|g_{\mu\nu}| \equiv g = 1.
\end{eqnarray} 
This condition must be preserved under coordinate transformations (diffeomorphisms). Hence, instead of an invariance under general coordinate transformation (general diffeomorphism) under which GR is based, UG is constructed by imposing that the constraint (\ref{umc}) is preserved, resulting in the invariance by transverse diffeomorphism \cite{trans}. Considering an infinitesimal coordinate transformation,
\begin{eqnarray}
x^\mu \rightarrow x^\mu. + \xi^\mu,
\end{eqnarray}
while in GR $\xi^\mu$ can be kept arbitrary, UG implies,
\begin{eqnarray}
\xi^\mu_{;\mu} = 0.
\end{eqnarray}

Condition (\ref{umc}) requires a very restrictive coordinate system in order to be satisfied. For example,  in a flat cosmological set, the determinant of the metric, using the lapse function $N$, is given by,
\begin{eqnarray}
g = Na^3.
\end{eqnarray}
Hence, (\ref{umc}) implies, $N = a^{-3}$. In this way, the time coordinate (let us call it $\theta$) is such that the metric is,
\begin{eqnarray}
ds^2. = a^{-6}d\tau^2 - a^2(dx^2 + dy^2 + dz^2).
\end{eqnarray}
In UG, many classical solutions (like the Schwarzschild one) may also be verified but using non-standard coordinate systems. This restriction can  be circumventend by changing (\ref{umc}) to,
\begin{eqnarray}
\label{umc-bis}
g = \chi,
\end{eqnarray}
where $\chi$ is an external function in the sense that it is non-dynamical but allowing, on the other hand, to use a convenient coordinate system.

UG can be implemented through an action as,
\begin{eqnarray}
{\cal L} = \sqrt{-g}\biggr\{R + \zeta(g - \chi)\biggl\} + {\cal L}_m,
\end{eqnarray}
where $\zeta$ is a Lagrangian multiplier and $\chi$ is an external field, as described above.
The resulting equations, after eliminating the Lagrangian multiplier, are
\begin{eqnarray}
\label{ue}
R_{\mu\nu} - \frac{1}{4}g_{\mu\nu}R = 8\pi G\biggr\{T_{\mu\nu} - \frac{1}{4}g_{\mu\nu}R\biggl\}.
\end{eqnarray}
These equations are traceless. In comparison to GR, this means that one information (precisely about the trace of the field equations) is lost.

Using the Bianchi identities in equations (\ref{ue}), we obtain,
\begin{eqnarray}
\label{nc}
\frac{R^{;\nu}}{4} + \frac{8\pi GT^{;\nu}}{4} = 8\pi G {T^{\mu\nu}}_{;\mu}.
\end{eqnarray}
Imposing the conservation of the energy-momentum tensor, the left-hand-side can be integrated, leading to,
\begin{eqnarray}
\frac{R}{4} + \frac{8\pi GT}{4} = \Lambda,
\end{eqnarray}
$\Lambda$ appearing as an integration constant. Hence, the final equations are,
\begin{eqnarray}
\label{ue-bis}
R_{\mu\nu} - \frac{1}{2}g_{\mu\nu}R &=& 8\pi G\,T_{\mu\nu} + g_{\mu\nu}\Lambda,\\
{T^{\mu\nu}}_{;\mu} &=& 0.
\end{eqnarray}
Hence, the UG equations become formally equivalent to the GR equations in presence of a cosmological constant.
However, it is important to remember that $\Lambda$ in (\ref{ue-bis}) appears as an integration constant which, in principle,
has no connection with the cosmological constant present in the Lovelock invariant, neither with a cosmological constant associated with a fluid representing the vacuum energy density. In this sense, it is stated generally that the UG at least alleviate the cosmological constant problem.

The previous statement can perhaps be made more clear by considering a self-interacting scalar field with an energy-momentum tensor given by,
\begin{eqnarray}
T_{\mu\nu} = \epsilon\biggr\{\phi_{;\mu}\phi_{;\nu} - \frac{1}{2}g_{\mu\nu}\biggl\} + g_{\mu\nu}V(\phi).
\end{eqnarray}
The parameter $\epsilon$ takes value $\pm 1$, representing an ordinary scalar field (positive sign) or a phantom field (negative sign).
The cosmological constant is given by the particular case $\phi =$ constant. In UG framework, and using the energy-momentum tensor for a self-interacting scalar field, the field equations are,
\begin{eqnarray}
\label{sea}
R_{\mu\nu} - \frac{1}{4}g_{\mu\nu}R &=& \epsilon\biggr(\phi_{;\mu}\phi_{;\nu} - \frac{1}{4}g_{\mu\nu}\phi_{;\rho}\phi^{;\rho}\biggl),\\
\label{seb}
\frac{R_{;\nu}}{4} &=& \epsilon\biggr(\phi_{;\nu}\Box\phi + \frac{\phi^{;\rho}\phi_{;\nu;\rho}}{2}\biggl).
\end{eqnarray}
Remark that the potential $V(\phi)$ has disappeared from the field equations. Moreover, the usual Klein-Gordon equation
has been replaced by a more complex structure.
If the usual Klein-Gordon equation
\begin{eqnarray}
\epsilon\Box\phi = - V_\phi
\end{eqnarray}
 is imposed (which is equivalent to impose the conservation of the energy-momentum tensor), equation (\ref{seb}), becomes,
 \begin{eqnarray}
 \label{sec}
\frac{R_{;\nu}}{4} &=& \epsilon\biggr(\phi_{;\nu}V_\phi + \frac{\phi^{;\rho}\phi_{;\nu;\rho}}{2}\biggl).
\end{eqnarray}
Equation (\ref{sec}) can now be integrated, leading to,
 \begin{eqnarray}
 \label{sec}
\frac{R}{4} &=& \epsilon\biggr(V(\phi) + \frac{\phi^{;\rho}\phi_{;\rho}}{4}\biggl) + \Lambda,
\end{eqnarray}
$\Lambda$ being again an integration constant.
The final equations are,
\begin{eqnarray}
\label{sea}
R_{\mu\nu} - \frac{1}{2}g_{\mu\nu}R &=& \epsilon\biggr(\phi_{;\mu}\phi_{;\nu} - \frac{1}{2}g_{\mu\nu}\phi_{;\rho}\phi^{;\rho}\biggl) + g_{\mu\nu}V(\phi) + g_{\mu\nu}\Lambda,\\
\label{seb}
\epsilon\Box\phi = - V_\phi.
\end{eqnarray}
Hence, $\Lambda$ has no (direct at least) relation with $V(\phi)$.

The considerations above show that UG may lead to a new vision on the presence of a cosmological constant in the gravitational equations. All these considerations rely on the conservation of the energy-momentum tensor. However, in UG the conservation of the energy-momentum tensor is not consequence of the theory itself as it happens in GR. In fact, GR is invariant by a general diffeomorphism and this lead to the conservation of the energy-momentum tensor \cite{wald}. The energy-momentum tensor is canonically defined from the matter Lagrangian ${\cal L}_m$ as,
\begin{eqnarray}
T_{\mu\nu} = - \frac{2}{\sqrt{-g}}\frac{\delta(\sqrt{-g}{\cal L}_m)}{\delta g^{\mu\nu}}.
\end{eqnarray}
Using the matter action,
\begin{eqnarray}
{\cal A}_m = \int d^4 x\sqrt{-g}{\cal L}_m,
\end{eqnarray}
it results,
\begin{eqnarray}
\delta_g{\cal A}_m = \delta_g\int d^4 x\sqrt{-g}{\cal L}_m = - \frac{1}{2}\int d^4x \sqrt{-g} T_{\mu\nu}\delta g^{\mu\nu},
\end{eqnarray}
If $\delta g^{\mu\nu} = \xi^{(\mu;\nu)}$, corresponding to a diffeomorphic transformation,
then
\begin{eqnarray}
\delta_\xi{\cal A}_m = - \frac{1}{2}\int d^4x \sqrt{-g} T_{\mu\nu} \xi^{\mu;\nu} = \frac{1}{2}\int d^4x \sqrt{-g} {T_{\mu\nu}}^{;\mu} \xi^\nu.
\end{eqnarray}
Hence, $\delta_\xi{\cal A}_m = 0$ implies the conservation of the energy-momentum tensor, ${T^{\mu\nu}}_{;\mu} = 0$.
However, UG is invariant by transverse diffeomorphism, and if
\begin{eqnarray}
{T^{\mu\nu}}_{;\mu} = \partial^\nu \Theta,
\end{eqnarray}
for a given scalar function $\Theta$, we have,
\begin{eqnarray}
\delta_\xi{\cal A}_m = \frac{1}{2}\int d^4x \sqrt{-g} {T_{\mu\nu}}^{;\mu} \xi^\nu = \frac{1}{2}\int d^4x \sqrt{-g} \Theta_{;\nu} \xi^\nu = \frac{1}{2}\int d^4x \sqrt{-g} \Theta \xi^\nu_{;\nu} = 0,
\end{eqnarray}
without the necessity to have the conservation of the energy-momentum tensor.

There are many discussions in the literature if such possibility of non-conservation of the energy-momentum tensor may be an extra tool to distinguish between GR and UG, at least at perturbative level, see \cite{pert,brand} and references therein.
Here, we would like to stress that, while the conservation of the energy-momentum tensor implies GR equations in presence of a cosmological constant, the non-conservation of the usual energy-momentum tensor leads to GR equations
with a dynamical cosmological term, which can be identified with a dynamical vacuum. This may be connected with interacting models.

Let us take as an example, the equation (\ref{nc}), and we identify,
\begin{eqnarray}
\label{ct}
\frac{R}{4} + \frac{8\pi GT}{4} = - \Lambda,
\end{eqnarray}
$\Lambda$ now being a non-constant function.
Hence,
\begin{eqnarray}
8\pi G {T^{\mu\nu}}_{;\mu} = - \Lambda^{;\nu}.
\end{eqnarray}
The final equations are,
\begin{eqnarray}
\label{ue-tri}
R_{\mu\nu} - \frac{1}{2}g_{\mu\nu}R = 8\pi G\,T_{\mu\nu} + g_{\mu\nu}\Lambda,\\
{T^{\mu\nu}}_{;\mu} =  - \Lambda^{;\nu}.
\end{eqnarray}
These equations are characteristic of a decaying vacuum theory, with an interaction of matter and the vacuum term.

All these considerations reveal the richness of the UG proposal, leading to many new perspectives to some problems that appear in GR, in particular the {\it cosmological constant problem}.

\section{Final remarks}

The search for alternative geometrical formulations for the description of the gravitational phenomena has begun almost at the same time General Relativity was proposed. By now, since GR has been proposed, more than one thousand alternative theories to GR have appeared, as Ruben Aldrovandi has emphasized in this speech in the {\it Laboratoire de Gravitation et Cosmologie Relativistes} about 30 years ago. The terminology has changed, today it is preferred to say {\it modified gravity} instead of {\it alternative theories of gravity}, but this profusion of possible new theories of gravity indicates perhaps some aspects of the study of the gravitational phenomena itself, and has a strong contrast with the fate of the newtonian theory, that was essentially the only theory of gravity until the emergence of GR.

The description of gravity as a geometric phenomena, linked with the structure of the space-time itself may be related to, at least, two aspects. The first, the universality of the gravity (absolutely everything is subjected to the gravitational interaction). Describing gravity as the structure of the space-time forcely implies this universality. Moreover, the appearance of the non-euclidean geometries about 200 years ago, resulting in an infinite numbers of possible, consistent, geometric structures, led to the question ``which geometrical structure nature has chosen?", a question that was senseless if just one geometry existed.  The GR answer to this was: ``Nature has not chosen any {\it a priori} geometry,  it is the matter distribution that creates dynamically the geometrical structure." 

Many new geometrical structures emerged, as the non-metricity, torsion, Finsler geometry, etc. Hence, in some sense is natural that many new possible theories of gravity appear, keeping the main cornerstone of the GR theory: Gravity is the structure of the space and time. On the other hand, the quoted remark by Aldrovandi, in my opinion, alert to two main 
dangers: to forget about the open possibilities still existing without giving up standard physics; the need of a clear motivation when proposing new, still untested, paths deviating from this standard physics. To explore all the possibilities 
existing in mathematics and physics is important, even crucial in order to open new windows in our comprehension of nature. However, it is important to keep in mind why some still unexplored ways are followed. Aldrovandi and collaborators have considered some important new variant, as the teleparelellism and de Sitter relativity. In this text, we have discussed two other traditional variant: the scalar-tensor gravity theories and the Unimodular Gravity.

The most paradigmatic scalar-tensor theory is the Brans-Dicke one. It implements an idea by Dirac about the possible variation of the
the gravitational coupling. Besides giving many different and new predictions compared with GR, the BD theory open an interesting new windows in the strong gravity regime, as it has occurred in the primordial universe, mainly concerning the Planck regime. In fact, a change in the gravitational coupling affects the Planck parameters, like energy, temperature, time and spatial scales. This may imply new important effects even if the BD parameter $\omega$ is huge as suggested by experiments and observations. In the present text, we used the BD theory to discuss, quite qualitatively, how a new vision of the dynamics of the cosmic, relating the expansion of the universe, as seen from one frame, to a non-expanding universe but with varying mass of the elementary particles. This may imply in some possible new cosmological scenarios, for example, concerning how to implement an inflationary era.

On the other side, Unimodular Gravity may give new perspectives to the cosmological constant problem. In some sense, UG may lead to {\it degravitation} of the vacuum energy, but implying at the end in a presence of a cosmological term in the field equations which may be not directly connected with vacuum energy. UG admits a modified energy-momentum tensor conservation law, with respect to the usual conservation law as it appears in GR, that opens new perspective. For example, it may lead naturally to a decaying vacuum state, with an interaction between ordinary matter and the fluid representing the vacuum. One important question is that if the final structure is equivalent to the corresponding structure in GR. Some previous investigation shows that this may happen at perturbative level, but this is not a closed issue. On the other hand, UG may lead to new results at quantum regime.

In spite of the proliferation of {\it deviations} of the Standard Theory of Gravity, as remarked by Aldrovandi, we think that the proposals discussed here try to keep contact with the concrete problems that physics faces when trying to give a more complete description of Nature, mainly when effects ordinarily originating from quantum field theory are applied to the domain of gravitation.

\bigskip

\noindent

{\bf Acknowledgements:} I thank CNPq and FAPES for partial financial support, and José Geraldo Pereira for inviting me to contribute to the book in honor of Ruben Aldrovandi. I thank Hermano Velten, Luiz Filipe Guimarães, Nelson Pinto Neto and Richard Kerner for their suggestions on the text.

 \end{document}